\documentclass[aps,prb,superscriptaddress,twocolumn,longbibliography]{revtex4-1}

\usepackage{lmodern}

\usepackage[latin9]{inputenc}
\usepackage{amsmath}
\usepackage{amssymb}
\usepackage{graphicx}
\usepackage{epstopdf}
\usepackage{color}
\usepackage{enumerate}
\usepackage{ulem}

\begin{document}
%\linenumbers

\title{Quantum breakdown of superconductivity in low-dimensional materials}

\author{Benjamin Sac\'{e}p\'{e}}
\affiliation{Univ. Grenoble Alpes, CNRS, Grenoble INP, Institut N\'{e}el, Grenoble, France}
\author{Mikhail Feigel'man}
\affiliation{L.D. Landau Institute for Theoretical Physics, Chernogolovka, Russia}
\affiliation{Skolkovo Institute for Science and Technology, Moscow, Russia}
\author{Teunis M.  Klapwijk}
\affiliation{Kavli Institute of Nanoscience, Delft University of Technology, Delft, The Netherlands}
\affiliation{Institute for Topological materials, Julius Maximilian University of Würzburg, Würzburg, Germany}
\affiliation{Physics Department, Moscow State University of Education, Moscow, Russia }

\begin{abstract}
In order to understand the emergence of superconductivity it is useful to study and identify the various pathways leading to the destruction of superconductivity. One way is to use the increase in Coulomb-repulsion due to the increase in disorder, which overpowers the attractive interaction responsible for Cooper-pair formation. A second pathway, applicable to uniformly disordered materials, is the competition between superconductivity and Anderson localization, which leads to electronic granularity in which phase and amplitude fluctuations of the superconducting order parameter play a role. Finally, a third pathway is an array of superconducting islands coupled by some form of proximity-effect, due to Andreev-reflections, and which leads from a superconducting state to a state with finite resistivity, which appears like a metallic groundstate. This review summarizes recent progress in understanding of these different pathways, including experiments in low dimensional materials and application in superconducting quantum devices.
\end{abstract}
\maketitle

\vspace{0.5cm}
\section{Introduction}

In parallel to the continuous discovery of new superconducting materials, a different research-strategy can be followed, which concentrates on the destruction of superconductivity in a given materials system. Without changing the composition a parameter, intrinsic to the material is changed such as electron density, disorder, or dimensionality, driving the transition from a superconducting state to another state, which is often found to be an insulator or a metal. In principle, this transition can be monitored at finite temperature ($T$), experimentally unavoidable, but the physics is focused on the transition which is expected to occur at zero temperature reflecting a transition from one ground state to another, a quantum phase transition~\cite{Sondhi97}. Such a quantum phase transition, with some theoretical bias traditionally called a superconductor-insulator transition~\cite{Goldman98}, has been studied in numerous experimental configurations and materials, encompassing amorphous thin films, granular superconductors, nanowires, gate-tunable superconducting oxide interfaces, aluminum-based Josephson junction arrays, proximitised metals and semiconductors or two-dimensional crystalline superconductors.

In the last decade, the actual appearance of the transition has revealed many surprises that have drastically changed our understanding of conventional superconductivity with some resonance with phenomena observed in high temperature cuprate superconductors. These new developments are the focus of this review article. To remove interpretative bias in the terminology we use in this review, instead of the commonly used term superconductor-insulator-transition (SIT), a more neutral term \textit{Quantum Breakdown of Superconductivity} (QBS), and use the term superconductor-insulator transition, when it is indeed a transition to an insulating state and superconductor-metal transition (SMT) if it transits into a metal-like state. As we will see even these 'insulator' and 'metal' states can also be based on ingredients related to the superconducting state, without showing zero resistance.     
 
%%%%%%%%%%%%%%%%%%%%%%%%%%%%%%%%%%%%%%%
%%%%%%%%%%%%%%%%%%%%%%%%%%%%%%%%%%%%%%%
%%%%%%%%%%%%%%%%%%%%%%%%%%%%%%%%%%%%%%%
\vspace{0.5cm}
\section{Main paradigms}
Various means can be employed to study experimentally the evolution of superconductivity towards the breakdown. Some obvious ones are the application of an external magnetic field ($B$) or a dc current, which leads to the transition to resistive superconducting states. Suitable materials can be created by either increasing disorder, changing the carrier density by field-effect gating or by using a lower effective dimensionality in thin-film, nanowire geometries or single atomic thick layers. Understanding how the  superconducting order parameter $\Psi = \Delta e^{i\phi} $ evolves, while the zero-resistance property is suppressed is the primary question that challenges several fundamental concepts and paradigms of condensed matter physics. 

The first key paradigm concerns the complex nature of the superconducting order parameter. The suppression of superconductivity can follow two main paths~\cite{Larkin99}. It is either a suppression of the amplitude $\Delta $, or a loss of the stiffness of the phase $\phi $. The former involves the interaction between the electrons --the fermionic channel-- which drives the strength of the attractive interaction leading to Cooper-pairing, and which may, if pushed from attractive to repulsive, restore the normal state~\cite{Finkelstein87,Finkelstein94}.  The second one  yields a less intuitive mechanism in which Cooper-pairing remains and resistive properties emerge because the macroscopic phase looses long-range order and varies in time~\cite{Fisher90}.  These two antagonistic mechanisms towards the destruction of superconductivity, have been dubbed the fermionic or bosonic scenario.  Both yield resistive states, terminating zero-resistance superconductivity, but very different  with respect to the nature of the charge carriers, either being single electrons or Cooper pairs.

Another important aspect to grasp is the effective dimensionality of the system. While for the single electron coherence the inelastic length is the relevant quantity in comparison to the thickness of a thin film or the width of a nanowire, for a superconductor the superconducting coherence length $\xi$ is the scale to compare with. Superconductivity is in the latter case quasi-two-(one-)dimensional when the thickness (or width) is smaller than $\xi$. This reduced dimensionality has major consequences for the superconducting state~\cite{Larkin05}: it enhances fluctuations and generates topological defects, vortices or phase slips, that drastically modify the transport properties and eventually suppress the phase stiffness through a Berezinskii-Kosterlitz-Thouless transition or through one-dimensional (quantum) phase slips.

Last but not least, disorder is an unavoidable  ingredient, which has deep consequences in limiting electron transport in general and which contrasts sharply with the zero-resistance property of superconductivity. Although disorder is at first sight incompatible with superconductivity, experiments and theory show that superconductivity develops in systems in which the single electrons are localized, leading to an insulating state in the  $T=0$ limit. If Cooper-pairing can develop with localized single-electron states, then the question arises whether one may encounter localization of Cooper-pairs. In the past decade, a large body of work has demonstrated that this type of localization occurs in some amorphous superconductors.  Probing and understanding the properties of insulating systems with localized Cooper-pairs poses further challenges and intriguing questions. 

The above rather generic ingredients, reduced dimensions, disorder, and Coulombic inter-electron interactions, can be made present in a controllable way in many superconductors. Through decades of research, primarily on  magnetotransport and using the analysis of quantum criticality it has been found that the breakdown of superconductivity does not follow a universal path.  Instead, there appear to be almost as many QBSs as systems under study.  This led to a classification of  various types by structural aspects, \textit{i.e.} granular \textit{versus} homogeneously disordered (amorphous) systems, by the level of charge carrier density, by their effective dimensionality, by the type of weak link in Josephson junction arrays, or by the driving parameter. In this review,  we have chosen to leave out a previously studied specific model-system of Josephson-tunnel-junction arrays with competing Josephson and charging energies~\cite{Fazio2001}. 

Our review aims at highlighting the new phenomena and concepts that emerged recently and led to revisit several longstanding paradigms in the field of the QBS. Experimentally, much of the recent progress is due to the use of very low temperature spectroscopy with local probes, unveiling the emergent electronic granularity in homogeneously disordered thin films and the existence of the pseudogap for preformed Cooper pairs. These discoveries led to the unexpected breakdown of the often cited fermionic and bosonic dichotomy and demanded  a new microscopic description of superconductivity subject to strong disorder. Moreover, in a number of experimental systems the transition to the superconducting state is found to be incomplete, that is, terminated by a metal-like state down to the lowest accessible temperature~\cite{KKS19}. This phenomenon might be called a superconductor-metal transition, although the origin of this vividly discussed~\cite{Tamir19,Dutta19} metallic state has not yet been unambiguously determined. We also discuss how new gate-tunable semiconductor and low-dimensional materials can be used to couple superconducting islands creating proximitized Josephson junction arrays, providing new insights into the QBS physics utilizing insights from mesoscopic physics. Finally we address how strong disorder modifies the electrodynamics near the QBS and how this electrodynamics can can serve in hybrid quantum circuits.

%%%%%%%%%%%%%%%%%%%%%%%%%%%%%%%%%%%%%%%
%%%%%%%%%%%%%%%%%%%%%%%%%%%%%%%%%%%%%%%
%%%%%%%%%%%%%%%%%%%%%%%%%%%%%%%%%%%%%%%
\section{Quantum Breakdown of Superconductivity: Amplitude \textit{vs} phase driven transition}

   \begin{figure*}[t!]
		\includegraphics[width=0.8\linewidth]{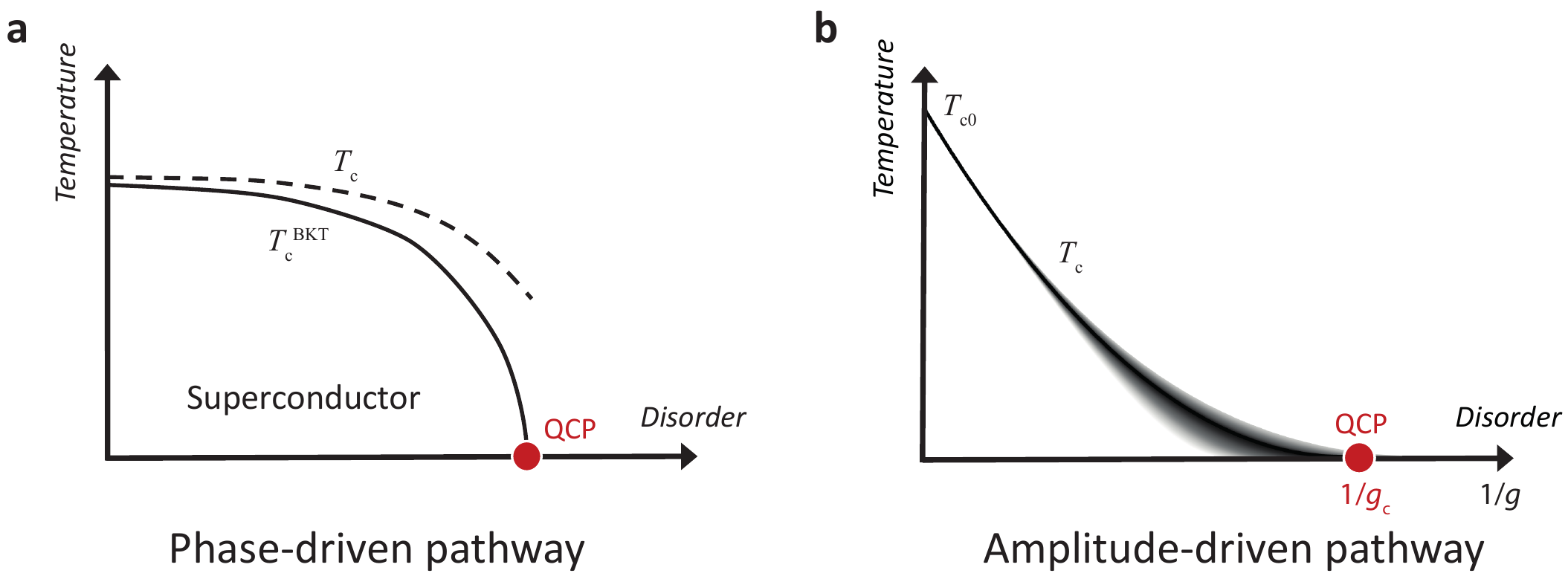} 		
   	\caption{\textbf{The phase diagrams of 2D superconductors.} \textbf{a,} Temperature-disorder phase diagram of 2D superconductors~\cite{Fisher90}. The transition to the superconducting state is defined by the critical temperature $T_c$ at which Cooper-pairs preforms. The superfluid stiffness, indicating superconductivity, develops at a lower temperature, below $T_c^{BKT}$, the critical temperature of the Berezinski-Kosterlitz-Thouless transition~\cite{Berezinskii71,Kosterlitz73}. On increasing disorder, $T_c$ is reduced by the Finkel'stein mechanism~\cite{Finkelstein87,Finkelstein94} illustrated in \textbf{b}. Similarly, $T_c^{BKT}$ is suppressed faster due to the disorder-enhanced phase fluctuations, till a critical disorder that defines the quantum critical point (QCP). The existence of the two critical temperatures opens up a sizeable temperature-regime for phase fluctuations~\cite{Emery1995} between $T_c^{BKT}$ and $T_c$, which grows with increasing disorder. At the quantum critical point the superfluid stiffness vanishes, but without the destruction of Cooper-pairs. Without stipulating the nature of the state terminating superconductivity, metal or insulator, one recognizes that this scenario defines a prototypical continuous quantum phase transition~\cite{Sondhi97}, driven by quantum fluctuations of the phase of the order parameter. \textbf{b,} Coulomb suppression of the critical temperature $T_c$ as a function of disorder $1/g$, with $g=h/e^2 R_{\square}$ according to Finkel'stein theory. The shaded grey area around the $T_c$ line represents the fluctuations of the local critical temperature $T_c(\mathbf{r})$ that develop and grow as $\frac{\delta T_c}{T_c} \approx  \frac{0.4}{g(g-g_c)}$ upon approaching the critical point  $1/g_c$ (Ref.~\cite{Skvortsov05}). Notice that the disorder-dependence of $T_c$ takes the form drawn in \textbf{a} when plotted as a function of $\log( 1/g)$, see for instance Fig. 4 and 5 in Ref.~\cite{Strongin70}.}
   	\label{Fig1}
   \end{figure*}
	
Thin superconducting films have become the prototypical systems to study the quantum breakdown of superconductivity. Due to their quasi-2-dimensional nature, thermal and quantum fluctuations of the order parameter $\Psi = \Delta e^{i\phi}$ play in thin films a crucial role~\cite{Larkin05,Emery1995}. They are furthermore enhanced by disorder that reduces the superconducting coherence length. At the same time, the diffusive motion of electron impeded by the reduced dimensionality tends to enhance the strength of the  Coulombic inter-electron interaction~\cite{Altshuler85b}. These two phenomena generate two distinct pathways to the QBS: either a suppression of the superconducting phase stiffness by quantum phase fluctuations, the phase-driven pathway, or a Coulomb-assisted suppression of the Cooper-pair attraction, the amplitude pathway.  

\subsection{Phase-driven pathway}
A seminal approach to the phase-driven QBS in thin-film superconductors was proposed by Matthew Fisher~\cite{Fisher90} who developed the quantum critical scaling theory of the dirty boson model. It is a 2D model of hard-core, interacting bosons of charge $2e$ (describing Cooper-pairs) in a random potential. The fundamental concept of this work, which motivated considerable experimental activity on thin superconducting films, is the possibility that, upon raising the magnetic field at $T=0$, the pinned vortices with increasing density, could delocalize and undergo a Bose condensation. Similarly for the vortex-anti-vortex glass phase upon raising disorder at $B=0$ (see Fig.~\ref{Fig1}a). This implies the localization of the charge $2e$-bosons mirroring the localization of vortices mandatory for the condensation of the charge $2e$-bosons in the superfluid phase. As a result the QBS in this model is the competition between condensation of Cooper pairs and of vortices. 

Two major outcomes of this theory have laid the groundwork for decades of work on this specific theory for the QBS. First, the resistance at the quantum critical point is predicted to be metallic due to the simultaneous presence of diffusion of vortices and charges, and should reach, assuming a self-duality between charges and vortices, the quantum resistance for charge $2e$, $h/{(2e)^2}$. Secondly, the quantum scaling analysis led to a frequently used scaling dependence of the resistance around the quantum critical point: 
\begin{equation}
R =\frac{h}{4e^2} \mathcal{F}\left( |\delta| / T^{1/z\nu} \right),
\label{eq3}
\end{equation}
which relates the behaviour of the resistance to the critical exponents  $\nu$ and $z$ of the diverging spatial and dynamical correlation lengths~\cite{Sondhi97}, via the scaling functional $\mathcal{F}$. The variable $\delta$  is defined as $\delta = (X-X_c)/X_c$, the distance to the critical point $X_c$ for the tuning parameter $X$. 

Experimentally, Hebard and Palaneen~\cite{Hebard90} were the first to uncover an intriguing crossing point in the magnetoresistance isotherms of amorphous InO (a:InO) films. The nearly $T$-independent resistance at the crossing point reached a value of $ 0.7 h/4e^2$ and was identified as the quantum critical point of the magnetic-field-driven SIT. Scaling analysis of the resistance data around it, taken at various $T$, were shown to collapse on a single functional, in accord with Eq.~(\ref{eq3}), providing direct access to the critical exponents $\nu$ and $z$. Another seminal work, conducted by Haviland et al.~\cite{Haviland89}, discovered a disorder (thickness)-tuned QBS in quench-condensed amorphous bismuth films and found that the critical temperature is continuously reduced upon increasing sheet resistance, reaching full suppression for a resistance of the order of $h/{(2e)^2}$. 

These experimental findings, together with Fisher's theory, lent support to a phase-driven QBS-scenario and stimulated a large body of experimental work, to establish the  universal character of the phase-driven QBS independent of the various systems and materials. The studies carried out over the past three decades repeatedly confirmed the presence of a, sometimes approximate, crossing point in the magnetoresistance and reproduced with varying degrees of success the quantum scaling of the data. Unfortunately, the resulting collection of critical exponents extracted from scaling analysis covers a large range of values, ranging from $z\nu =0.6$ to $2.4$. In order to reconcile this dispersion, different universality classes were invoked, for example for classical and quantum percolation~\cite{Steiner08} (see for example the reviews~\cite{Gantmakher10,Lin15}). Even then, one finds a lack of universality for the critical exponents, an often very limited range of temperature or field usable in the scaling analysis, a non-perfect crossing point that sometimes transforms into multiple crossing points in different temperature ranges, which is then interpreted as multiple quantum criticality~\cite{Biscaras13}. All these observations, have somehow undermined the confidence in arguments based on the data scaling analysis. In retrospect, the quantum scaling analysis, which continues to be widely used as an indicator for quantum criticality has become a mandatory figure for a phase-driven QBS while, at the same time, it has not been able to bring fruitful insight into the QBS physics, such as for example the understanding of the microscopic origin of the magnetoresistance crossing point.

Likewise, the critical resistance  is found to cover a range from $1$ to $30\,k\Omega$, although with a tendency to fluctuate from sample to sample around the resistance quantum $h/{(2e)^2}$ for strongly disordered systems~\cite{Murthy06}. However, sometimes with a remarkable accuracy such as for instance in the high-$T_c$ cuprate thin films~\cite{Bollinger11} and some graphene tin-decorated hybrid devices~\cite{Allain12}. But it can also approach $h/e^2$ as observed in the disorder-tuned QBS in TiN thin films~\cite{Baturina07} without any metallic separatrix between the superconducting and insulating films, contrary to what is expected for the dirty boson model. Consequently, such a diversity of critical resistance values points towards the need to consider carefully  other ingredients than just long-wave length phase fluctuations. These additional ingredients may be system-specific, which implies less emphasis on a conjectured universal behavior,  and more on the microscopics of the models for specific system-classes.

\subsection{Amplitude-driven pathway} 
The second major pathway to the quantum breakdown of superconductivity is based on the enhancement of the effective Coulomb repulsion due to the decrease in the diffusive motion of electrons~\cite{Altshuler85b}. Such a disorder--driven enhancement of the Coulomb interaction competes with the phonon-mediated attractive part of the interaction in the Cooper channel. The resulting continuous reduction of the effective attractive interaction leads to an amplitude-driven QBS, in practice a superconductor-metal transition with a vanishing pairing amplitude at the critical disorder.  First calculated with a perturbative diagrammatic technique~\cite{Maekawa82,Takagi82},  the full dependence of $T_c(R_{\square})$ was obtained with the renormalization group method by Finkel'stein~\cite{Finkelstein87,Finkelstein94}, which subsequently provided a simple analytical prediction for the critical disorder: $g_c={1}/{2\pi} \left(  \ln({1}/{T_{c0}\tau}) \right)^2$, with $g=h/(e^2 R_{\square})$, the dimensionless sheet conductance. Remarkably, $g_c$ is defined entirely by only two parameters: the unsuppressed $T_{c0}$ and the elastic scattering time $\tau$. The typical dependence of $T_c$ on disorder is illustrated in Fig.~\ref{Fig1}b.

The Finkel'stein theory has proven to describe successfully the suppression of $T_c$ in some thin films with low critical disorder~\cite{Finkelstein94}, meaning $g_c \gg 4$. In this case the state found after terminating superconductivity is a bad metal subject to weak localization effects at the experimentally accessible temperatures. Its intrinsic mechanism, which is just an extension of mean-field Bardeen-Cooper-Schrieffer (BCS) theory including disorder-enhanced interaction, is expected to be somewhat universal. However, on a quantitative level, many systems exhibit a critical disorder beyond the range of applicability of the Finkel'stein theory ($R_{\square} \sim h/4e^2$), where phase fluctuations are expected to contribute significantly~\cite{Larkin99}. When the sheet resistance reaches the resistance quantum, quantum phase fluctuations, localization effects and disorder-induced spatial inhomogeneities of the electronic properties disturb the standard dichotomy between phase and amplitude-driven pathways, leading to new intertwined scenarios.

%%%%%%%%%%%%%%%%%%%%%%%%%%%%%%%%%%%%%%%
%%%%%%%%%%%%%%%%%%%%%%%%%%%%%%%%%%%%%%%
%%%%%%%%%%%%%%%%%%%%%%%%%%%%%%%%%%%%%%%
\vspace{0.5cm}
\section{Emergent granularity of superconductivity at the local scale}

In order to create some insight in the diversity of experimental data on QBS a distinction was made between granular and homogeneous systems. The granular systems were usually mapped on SIS-type Josephson tunnel-junction arrays, following a phase-driven pathway to the breakdown of superconductivity. The homogeneous systems, were expected to show in its atomic structure some form of "homogeneous" (short-range correlated) disorder, which was expected to generate also some homogeneous superconducting state, subject to Coulomb-suppression of superconductivity. The last decade has unveiled a different, more complex and rich situation. Disorder, in the strong scattering limit when the mean-free-path is of the order of the interatomic distance, showed up as a strong disturbing agent that generates, unexpectedly, strong spatial fluctuations of the superconductivity-related spectral properties, enhanced by the proximity to the critical disorder~\cite{Sacepe08,Sacepe10, Mondal11,Sacepe11,Chand12,Sherman12, Noat13,Ganguly17,KunZhao2019}. The previously introduced classification between homogeneous and granular disordered materials became apparently  problematic, because homogeneously disordered materials showed self-induced electronic inhomogeneities, that is, emergent granularity of superconductivity, without an evident correlation with any structural granularity. 

%%%%%%%%%%%%%%%%%%%%%%%%%%%%%%%%%
\subsection{Emergent superconducting granularity}
   \begin{figure*}[t!]
	  \includegraphics[width=0.8\linewidth]{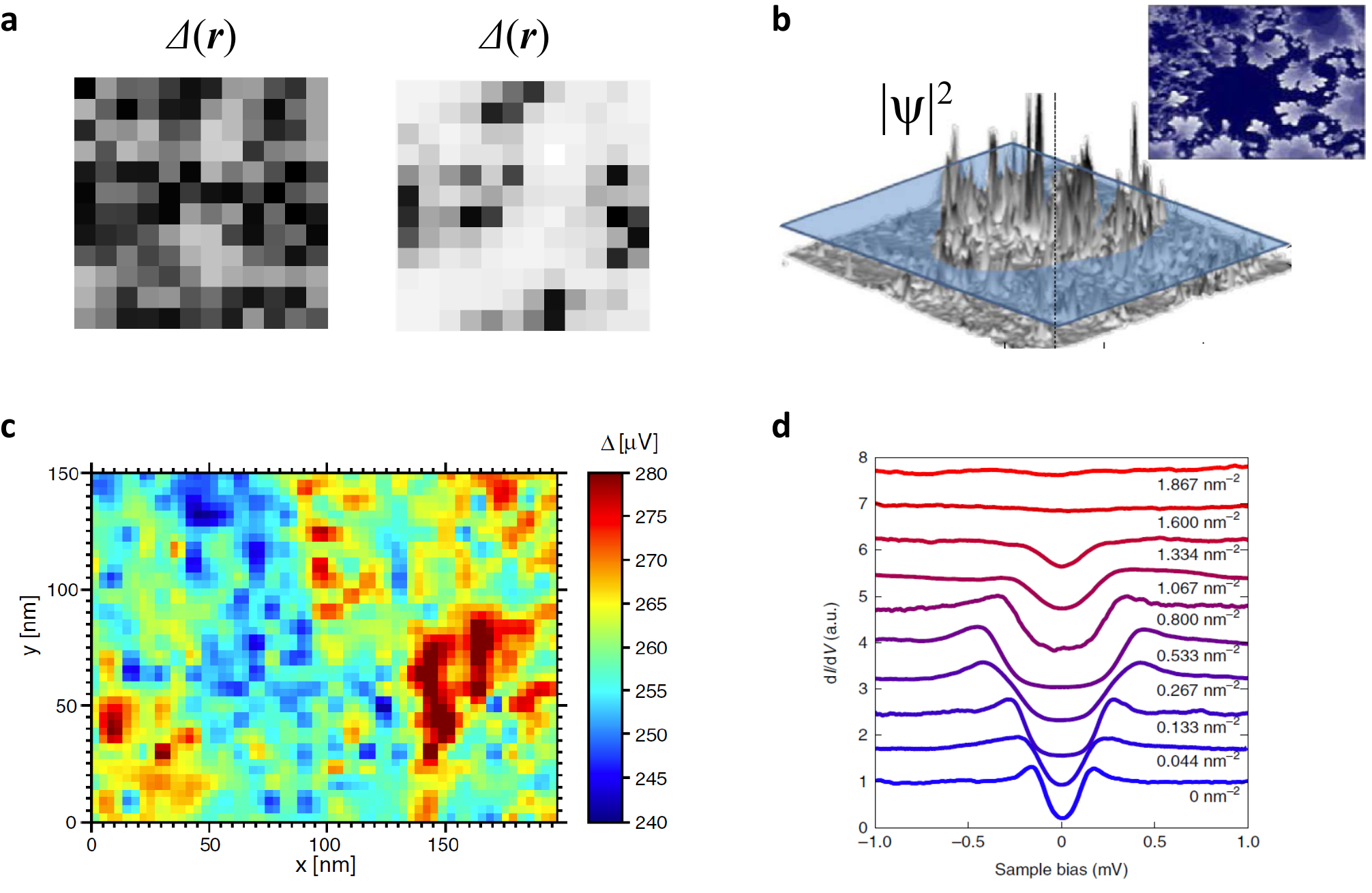} 		
   	\caption{\textbf{Emergent superconducting granularity.} \textbf{a,} Spatial map of the pairing amplitude $\Delta (\mathbf{r})$ obtained by numerical solution of the 2D disordered attractive Hubbard model~\cite{Ghosal98} with a disorder level equal to the near-neighbor hopping (left) and to twice the near-neighbor hopping (right). Sites with darker gray-scale indicate larger $\Delta (\mathbf{r})$'s. \textbf{b,} Fractal wavefunction intensity $|\Phi|^2$ at the mobility edge for the Anderson problem. The fractal nature is readily seen in the inset that shows the spatial wavefunction distribution at the intensity indicated by the blue plane. The wavefunction occupies only a fraction of the available volume. \textbf{c,} Spatial map of the superconducting gap measured by scanning tunneling spectroscopy on a TiN thin film~\cite{Sacepe08} near the QBS ($T_c \simeq 0.3\, T_{c0}$ and $R_{square} = 3.5\,k\Omega$ reached before the superconducting transition). Measurements were performed at $0.05$~K. \textbf{d,} Tunneling conductance $\text{d}I/\text{d}V$ measured by scanning tunneling spectroscopy on an epitaxial monolayer of NbSe$_2$ covered by Si adatoms~\cite{KunZhao2019}. Each spectrum corresponds to a different surface density of adatoms, that is, different level of disorder. The superconducting gap evolves non-monotonously with the surface density of adatoms. Reproduced from ref.~\cite{Ghosal98}, APS (\textbf{a}); ref.~\cite{Sacepe08}, APS (\textbf{c}); ref.~\cite{KunZhao2019}, (\textbf{d}). (\textbf{b}) courtesy of V. Kravtsov. }
   	\label{Fig2}
   \end{figure*}
The concept of disorder-induced inhomogeneities of the superconducting state was proposed back in 1971 by Larkin and Ovchinnikov~\cite{Larkin72}. It was understood that relatively small spatial variations in the Cooper attraction strength $\lambda(\mathbf{r})$  induced by disorder may lead to strong fluctuations of the local "transition temperature" $T_c(\mathbf{r})$. Since $T_c(\mathbf{r}) \propto \exp\left(-1/\lambda(\mathbf{r})\right)$,  the variation in $T_c$ is given by  $\delta T_c(\mathbf{r})/T_c  \sim \delta\lambda(\mathbf{r})/{\bar{\lambda}}^2$ with the average $\bar{\lambda} \ll 1$. These possible fluctuations and their impact on the superconducting properties have attracted considerable theoretical attention recently~\cite{Ghosal98,Meyer01,Ghosal01,Skvortsov05,Dubi07,Bouadim11,Feigelman12,Lemarie13,Evers2019}.  

The concept of spatial inhomogeneity of superconductivity was also analyzed later on within a different theoretical approach initiated by Ma and Lee~\cite{Ma85}. Assuming that, for some reason, the Finkel'stein mechanism is ineffective, the destruction of superconductivity is considered to be the result of Anderson localization~\cite{Ma85,Kapitulnik85,Kotliar86,Ghosal98,Ghosal01,Feigelman07,Feigelman10a}. Earlier work~\cite{Ghosal98,Ghosal01} demonstrated, by solving numerically the self-consistent Bogoliubov-De Gennes equations, that in the presence of significant local disorder of the on-site electron energies, the order parameter defined as the quantum-statistical average, $\Delta_{op}(\mathbf{r}) = \lambda \langle \psi_{\uparrow}(\mathbf(r)\psi_{\downarrow}(\mathbf{r})\rangle $, becomes strongly inhomogeneous in space. Fig.~\ref{Fig2}a shows the resulting fluctuations for the pairing amplitude calculated by Ghosal \textit{et al.}~\cite{Ghosal98}. Furthermore, the electron excitation spectrum of this inhomogeneous superconducting state is found to exhibit unusual features: the spectral gap does not coincide with the order parameter, as it does for the standard BCS superconducting state, and the gap edge singularities (coherence peaks) in the density of states become smeared. The spatial fluctuations of the coherence peak height provides another indicator characteristic of the superconducting inhomogeneities.

The numerical simulations~\cite{Ghosal98,Ghosal01} demonstrated that sufficiently strong disorder suppresses the coherence peaks completely, while keeping intact the spectral gap, thus opening an important question on the nature of the state terminating superconductivity. More recent numerical work~\cite{Dubi07,Bouadim11,Lemarie13}, some of them based on the quantum Monte Carlo method~\cite{Bouadim11}, which takes into account the quantum phase fluctuations between the self-induced superconducting islands, showed that the ground state is an insulator with a spectral gap caused by the attractive interaction. This type of  electron pairing is similar to the so-called "negative-$U$ Hubbard model" introduced by Anderson~\cite{Anderson76}. 

The body of numerical work reviewed above demonstrates the concept of disorder-induced "emergent superconducting granularity" and predicts new spectral features that contrast with the weak-disorder BCS superconductivity. However, numerical approaches are limited to small system-sizes and to the strong coupling limit, whereas superconductivity is in essence a "weak effect". Very recent results~\cite{Evers2019}  indicate a way to solve this problem.

%%%%%%%%%%%%%%%%%%%%%%%%%%%%%%%%%
\subsection{Superconductivity near the mobility edge}
The existing theory of weak-coupling superconductivity has been extended into the strong-disorder regime, by a combined analytical and numerical approach~\cite{Feigelman07,Feigelman10a}. The crucial aspect of this new theoretical development lies in the fractal nature of the nearly-localized electron wavefunctions when the Fermi level $E_F$ is close to the Anderson mobility edge $E_c$ (see Ref.~\cite{Kravtsov07}). These fractal electron wavefunctions $\psi_i(\mathbf{r})$ feature many unusual properties, such as,  intensities  $\psi_i^2(\mathbf{r})$ that fluctuate spatially very strongly (see Fig.~\ref{Fig2}b), multifractal statistics~\cite{Evers08}, and matrix elements $M_{ij} = \int dr \psi_i^2(\mathbf{r})\psi_j^2(\mathbf{r})$ that grow as $\left(E_c/\omega\right)^{\gamma}$ at a small energy difference $\omega \equiv |\xi_i - \xi_j|$. 
In addition, right at $E_F=E_c$ the inverse participation ratio $P_i = \int dr \psi_i^4(\mathbf{r})$ scales with the system size $L$ as $P \propto L^{-d_2}$, where $d_2 $ is the fractal dimension related to the exponent $\gamma = 1 - d_2/d$. Note that the three-dimensional Anderson mobility edge is characterized by $d_2 \approx 1.3$ and $\gamma \approx 0.6$, see Ref. [\onlinecite{Evers08}]. Importantly, fractal properties of wavefunctions are very robust at strong disorder, even half-way the mobility edge~\cite{Kravtsov07}.

These unusual properties profoundly modify superconductivity and have led to the concept of \textit{fractal superconductivity}. Using a generalized mean-field theory, Feigel'man \textit{et al.}~\cite{Feigelman10a} showed that the power-law scaling of $M(\omega)$ leads to a new dependence of the critical temperature on the microscopic parameters, which reads:
\begin{equation}
T_{c}(\lambda) \sim E_F\lambda^{1/\gamma}.
\label{Tc0}
\end{equation}
 This equation leads to the unexpected prediction that, for a constant $\lambda $, the critical temperature increases upon approaching the mobility edge compared to its weak disorder value. Later the same effect of $T_c$-enhancement due to fractality  was addressed by means of the renormalization group approach~\cite{Burmistrov12,Burmistrov15} in the 2D limit. It was shown that the partial suppression of the (screened) Coulomb amplitude makes the enhancement of $T_c$ by disorder even possible in two dimensions.

The results of Refs.~\cite{Feigelman07,Feigelman10a,Burmistrov12,Burmistrov15} were obtained within generalized mean-field approximation that neglects thermal phase fluctuations~\cite{Emery1995}. However, it was shown in Ref.~\cite{Feigelman10a} that the intensity of phase fluctuations is only moderate for superconductivity at the mobility edge, so the major conclusion on $T_c$ enhancement is valid. On the other hand, phase fluctuations do become crucial and destroy superconductivity when the Fermi level moves sufficiently far into the Anderson-localized band.

Another important feature of superconductivity near the mobility edge is the highly inhomogeneous superconducting order parameter. The dispersion $\sqrt{\langle \Delta^2(\mathbf{r}) \rangle}$ of the order parameter $\Delta(\mathbf{r})$ is much larger~\cite{Feigelman10a} than its mean value $\langle \Delta(\mathbf{r}) \rangle$, leading to a picture of a superconducting state splintered into superconducting islands. Such a strongly inhomogeneous superconductivity is a direct consequence of the fact that fractal wavefunctions occupy only a small fraction of the available volume (see Fig.~\ref{Fig2}b). A direct consequence of the inhomogeneity of $\Delta(\mathbf{r})$ is a strong vortex pinning and measurable critical currents even extremely close to the upper critical field~\cite{Sacepe19}.

%%%%%%%%%%%%%%%%%%%%%%%%%%%%%%%%%
\subsection{Emergent granularity within the amplitude pathway}
The theoretical approaches reviewed above predicted the emergence of an inhomogeneous superconducting state in the strong disorder limit. These theories however do not include the unavoidable Coulomb interaction that affects the attractive coupling via Finkel'stein's mechanism. Solving the combined effect of Anderson localization and Coulomb interaction  is known to be a notoriously difficult and still outstanding task. Nevertheless, some progress~\cite{Skvortsov05} was made on the perturbative level by including quantum interference effects,  the universal conductance fluctuations to the Finkel'stein's theory of the SMT.  Such an extended Finkel'stein theory predicts significant spatial fluctuations of the local transition temperature on approaching the quantum critical point of the theory, $g_c$, due to the disorder-enhancement of the mesoscopic fluctuations of the effective Coulomb amplitude. The fluctuations of $T_c$ are given by $\frac{\delta T_c}{T_c} \approx  \frac{0.4}{g(g-g_c)}$, which shows that these fluctuations necessarily become strong upon approaching the quantum critical point as illustrated by the shaded region around the mean $T_c$ in Fig.~\ref{Fig1}b.

%%%%%%%%%%%%%%%%%%%%%%%%%%%%%%%%%
\subsection{Intermediate conclusion regarding the theory}
The major outcome of these recent theoretical approaches is a clear breakdown of the conventional dichotomy between amplitude and phase-driven pathways. An initial normal state with homogeneous electronic properties can yield an inhomogeneous superconducting state whose spatial fluctuations get enhanced in the vicinity of the critical disorder. Such a formation of superconducting islands immersed in a non-superconducting matrix that would be metallic in the Finkel'stein SMT scenario and insulating in the strong disorder limit, intimately involves phase fluctuations between the weakly coupled superconducting islands. A new scenario emerges now and involves a remarkably complex and subtle interplay between localization phenomena (effect of disorder and multifractality), Coulomb interaction and phase fluctuations. The breadth of consequences of these superconducting inhomogeneities on the transport and thermodynamic properties is not yet understood and will certainly produce new conceptual advances. In addition, it makes the system sensitive to external conditions in a variety of experiments. 
 
We cite a few more theoretical consequences.  The spontaneous formation of superconducting islands leads to an enhancement~\cite{Skvortsov04} of the electron dephasing rate $1/\tau_\phi$  at low temperatures. Andreev reflections of electrons that develop between the superconducting islands (with uncorrelated fluctuations of their phases) become a dominant mechanism of decoherence~\cite{Skvortsov04}, far exceeding the usual Coulomb contribution $1/\tau_{\phi}^C \sim T/g$. This has implication for the possible existence of some anomalous metallic phases~\cite{KKS19}. Under perpendicular magnetic field, the self-induced inhomogeneities are predicted to induce multiple reentrant superconducting phases above the upper critical field in mesoscopic samples~\cite{SpivakZhou95}. More generally, disorder-induced inhomogeneities leads~\cite{TikhonovFeigelman2020} to the breakdown of the scaling theory~\cite{Belitz97} of the quantum SMT, which states that large-scale superconducting fluctuations are irrelevant due to long-range of proximity coupling  via the metal matrix.

%%%%%%%%%%%%%%%%%%%%%%%%%%%%%%%%%
\subsection{Real-space visualization of superconducting inhomogeneities}
In experiments, a continuous reduction of $T_c$ with increasing sheet resistance, consistent with Finkel'stein's theory, has long been the hallmark of structurally homogenous materials~\cite{Haviland89,Frydman03}, and, consequently, assumed  homogeneous superconductivity. With the progress in low temperature scanning tunneling microscopy, the local density-of-states can be measured with sub-kelvin resolution. The tunneling experiments have shed new light on this assumed homogeneity, in practice revealing an emergent granularity of  superconductivity on the local scale.

In 2008, Sac\'{e}p\'{e} \textit{et al.}~\cite{Sacepe08} reported the first scanning tunneling spectroscopy of thin superconducting films near the QBS. They studied TiN films, which exhibit a QBS~\cite{Baturina07} with a continuous reduction of $T_c$ and a high critical disorder of the order of $30\,k\Omega$.  The spectra of the local tunnelling density-of-states revealed significant spatial fluctuations of the superconducting gap $ \Delta (\mathbf{r}) $ on the scale of tens of nanometers. Fig.~\ref{Fig2}c shows a typical spatial map of $ \Delta (\mathbf{r}) $ measured at $0.05~K$. Upon approaching the critical disorder, these fluctuations were shown to increase from $ \delta\Delta  (\mathbf{r})/\Delta \simeq 0.15$ ($\Delta $ is here the average superconducting gap) for an intermediate disorder ($T_c \simeq 0.3\, T_{c0}$) up to $ \delta\Delta  (\mathbf{r})/\Delta\simeq 0.5$ for a nearly critical sample with $T_c \simeq 0.1\, T_{c0}$. Similar results were obtained in a series of studies by Raychaudhuri and co-workers on thick NbN films~\cite{Chand12,Lemarie13} and by Roditchev and co-workers~\cite{Noat13,Carbillet16,Carbillet19} on ultra-thin NbN films where the initial suppression of $T_c$ with disorder also seems to follow the amplitude pathway.  Cabrillet \textit{et al.}~\cite{Carbillet16} demonstrated by combining topography and scanning tunneling spectroscopy data the absence of any spatial correlation between small-scale structural grains and larger-scale fluctuations of the gap $\Delta(\mathbf{r})$. In a more recent work, Cabrillet \textit{et al.}~\cite{Carbillet19} also showed a clear anti-correlation between the width of the gap and the slope of the high-voltage anomaly in the tunnelling conductance, attributed to the "soft Coulomb gap"~\cite{AAL80}. This anti-correlation can be accounted for by the understanding that regions with larger local resistance are expected to have smaller $\Delta(\mathbf{r})$ due to the disorder-enhanced Coulomb effects~\cite{Finkelstein87,Finkelstein94}. Measurements in magnetic field add an additional perspective to the above picture.  Ganguly et al.~\cite{Ganguly17}  showed that NbN films not close to the QBS and exhibiting uniform superconducting properties at $B=0$ develop strong spatial inhomogeneities under perpendicular magnetic field values of 4 to 7.5~T, in agreement with theory~\cite{SpivakZhou95,Galitski01}. 

In general, the experiments on both TiN~\cite{Sacepe08} and NbN~\cite{Lemarie13,Noat13,Carbillet16,Ganguly17,Carbillet19} films demonstrate the same trend. An increase of the sheet resistance close to $R_Q=h/4e^2$ and a related suppression of $T_c$ are systematically accompanied by an increase of the gap-fluctuations $\Delta(\mathbf{r})$. Furthermore, the coherence peaks also fluctuate spatially and provide a measure of the amplitude of the local order parameter~\cite{Sacepe11,Lemarie13}.

The disorder-induced enhancement of the superconducting gap inhomogeneities is accompanied by a significant increase of the ratio $\Delta /T_c$ in TiN films~\cite{Sacepe08}, NbN~\cite{Mondal11} and in MoGe~\cite{Lotnyk17},  upon approaching the critical disorder. It grows in TiN films much above the weak-coupling value of $1.76$, up to $4$, indicating a serious deviation from the standard ratio of the BCS theory. The same evolution is seen also in thick, much thicker than the superconducting coherence length, a:InO films~\cite{Sacepe11,Dubouchet19}. In this case  the ratio of the spectral gap to $T_c$ grows from $2.5$ to $5.5$ when $T_c$ is reduced from $3.5$~K to $1.2$~K with increased disorder.  

A remarkable consequence of the increase of $\Delta /T_c$ with disorder is the non-vanishing spectral gap in nearly critical samples in spite of the enhanced spatial fluctuations. The anomalously large and increasing $\Delta /T_c$  ratio indicates that, at the critical disorder defined by $T_c=0$, the spectral gap remains finite and potentially persists into the insulator phase~\cite{Sacepe08,Sherman12}. Such behaviour would be consistent with the prediction of a gapped insulating phase (see Fig.~\ref{Fig3}d) in the disordered attractive Hubbard model~\cite{Ghosal98,Ghosal01,Bouadim11} discussed in the previous section. Furthermore, locations with a vanishing local order parameter, that is, without coherence peaks, remains fully gapped, as in a:InO~\cite{Sacepe11}, or with the presence of sub-gap states~\cite{Sacepe08,Chand12,Lemarie13,Noat13,Szabo16,Carbillet19}. The origin of the latter in nearly critical films with very low $T_c$ remains unclear and can be the result of pair-breaking due to interactions~\cite{SkvortsovFeigelman2013}, or more trivially limited by the energetic resolution of tunneling spectroscopy, which is notoriously known to be sensitive to the filtering of the electromagnetic environment~\cite{leSueur06,Martinis93}.

Recently, Zhao \textit{et al.}~\cite{KunZhao2019} investigated by low temperature scanning tunneling spectroscopy epitaxial monolayer of NbSe$_2$ on which disorder was controlled through \textit{in-situ} adatom deposition  prior to tunneling measurements. They observed an initial increase of the superconducting gap with disorder, followed by a sharp drop, as shown in Fig.~\ref{Fig2}d.  It may possibly constitute evidence for enhancement of 2D superconductivity by disorder~\cite{Burmistrov12}, although additional checks for alternative mechanisms are needed.

These series of scanning tunneling experiments provided compelling evidence for the emergent superconducting granularity in homogeneously disordered materials. The new picture of the superconducting state in the vicinity of the QBS is that of superconducting puddles embedded in a matrix with vanishing local order parameter, a matrix that can be gapped or gapless in case of a SIT or SMT, respectively. Consequently, quantum phase fluctuations in such weakly coupled puddles will definitely be crucial in the ultimate suppression of superconductivity. These superconducting inhomogeneities that emerge near criticality lead us to conclude to a progressive evolution from amplitude to phase-driven mechanism of QBS upon increasing disorder, at least in the materials investigated by scanning tunneling spectroscopy, until now: TiN, NbN and a:InO.

%%%%%%%%%%%%%%%%%%%%%%%%%%%%%%%%%%%%
%%%%%%%%%%%%%%%%%%%%%%%%%%%%%%%%%%%%
%%%%%%%%%%%%%%%%%%%%%%%%%%%%%%%%%%%%%
\section{Preformed pairs and their localization at the QBS}

In strongly disordered superconductors one can not \textit{a priori} assume that the attractive interaction leading to Cooper-pair formation is uniformly spread through the material. One possibility is that the attractive interaction exists only at randomly distributed local sites and that the Cooper-pairs are also formed locally. Since, such a case does not necessarily lead to long range coherence, these pairs are called \textit{preformed pairs}, anticipating the development of a state of superconductivity by phase-coherence between these localised pairs. Such a theoretical possibility, which is obviously difficult to identify based on transport measurements, has recently received experimental support from local spectroscopy data.  A suppression of the quasiparticle tunneling density of states was observed at temperatures far above the superconducting transition temperature, in strongly disordered TiN thin films~\cite{Sacepe10}. A similar suppression is routinely observed in underdoped cuprate superconductors and labeled a \textit{pseudo-gap}~\cite{Fischer07}.  Assuming that this pseudo-gap (PG) is related to superconductivity, one is led to conclude that on cooling down initially \textit{preformed pairs} of electrons appear, which become coherent at much lower temperatures, at the critical temperature $T_c$. An early analysis of the tunnel-current above $T_c$ in the regime of superconducting fluctuations was carried out by Varlamov and Dorin~\cite{Varlamov83}, signalling theoretically an apparent  pseudo-gap formation.

\begin{figure*}[t!]
   	\includegraphics[width=1\linewidth]{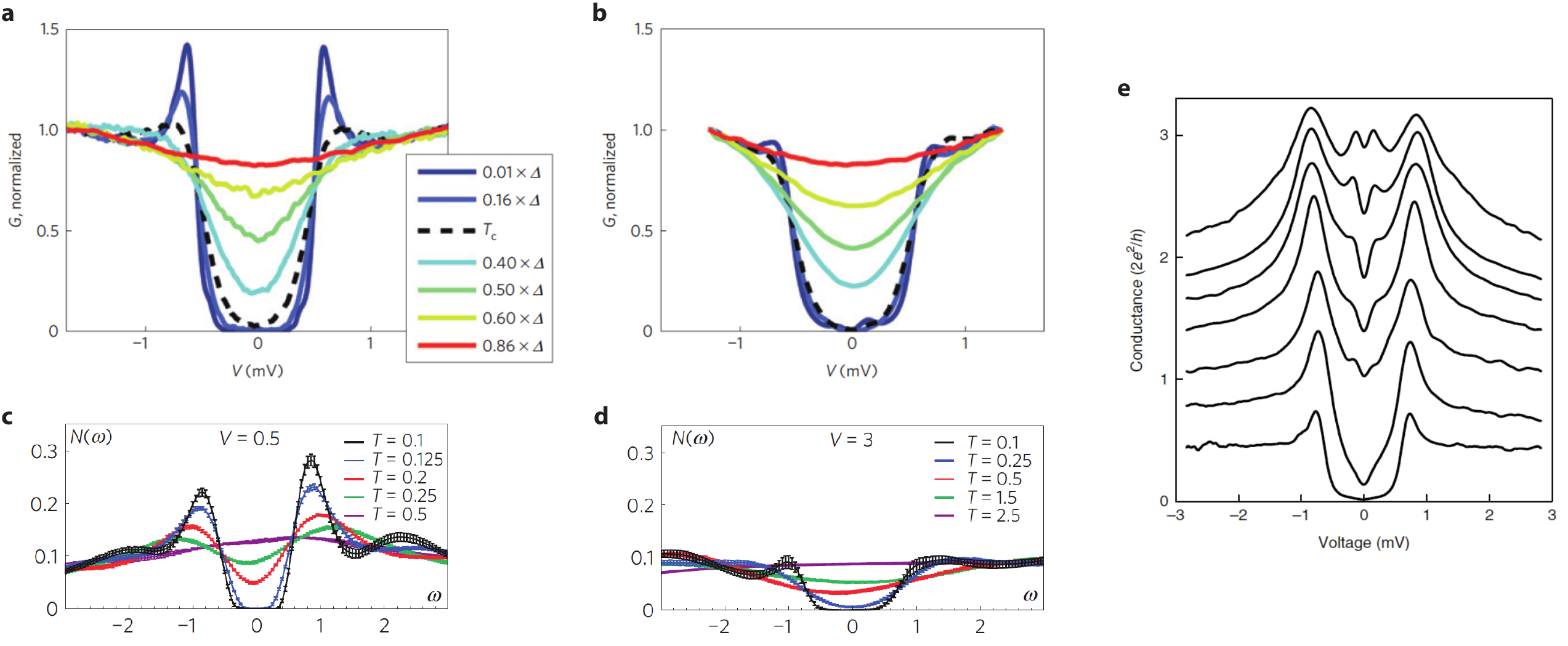} 		
\caption{\textbf{Pseudogap and collective gap of preformed Cooper pairs.} \textbf{a} and \textit{b,}Temperature-evolution of the local tunneling conductance $G$ versus voltage bias $V$, characterized by the presence (a) or absence (b) of superconducting coherent peaks~\cite{Sacepe11}. Both sets of data were measured, at a two different locations in superconducting samples. The tunneling spectra are selected at temperatures equal to fractions of the low-$T$ spectral gap. The spectral gap values are $\Delta = 560\,\mu$eV and $\Delta = 500\,\mu$eV for \textbf{a} and \textbf{b} respectively. The black dashed lines show the spectra measured at $T_c$. The clear pseudogap without coherence peaks and without state at the Fermi level ($V=0$) is the signature of preformed Cooper-pairs. Both sets of data are representative of superconducting samples with emergent granularity, for which superconducting islands show spectra with coherence peaks (a) and the surrounding matrix a gap without coherence peaks (b). The latter being a spectral signature of localized preformed Cooper pairs~\cite{Sacepe11}. \textbf{c,} Numerical simulations of the disorder-averaged density-of-states $N(\omega)$ for the 2D disordered attractive Hubbard model. A pseudogap develops for $T\simeq T_c=0.14$ and the coherence peaks vanish at $T\sim T_c$, in remarkable agreement with the experimental data in (\textit{a}). \textbf{d,} Similar simulations but for the insulating state at high disorder. The temperature-evolution of $N(\omega)$ resembles that of  the superconducting case in \textbf{c} but without coherence peaks. This set of simulations indicate that the insulator is gapped due to electron pairing. \textbf{e,} Point contact Andreev spectroscopy: evolution of the local differential conductance $G = \text{d}I/\text{d}V$ versus bias voltage measured on an a:InO sample at $T = 0.065$~K and at the same position for different values of the point-contact conductance. The conductance curves are normalized to $2e^2/h$ and have not been vertically shifted. The evolution from the tunneling to the Andreev spectroscopy unveils a new pair of peaks inside the single-electron gap, which relate to the collective gap  $\Delta_{\mathrm {col}}$. Reproduced from ref.~\cite{Sacepe11}, (\textbf{a,b}); ref.~\cite{Bouadim11}, (\textbf{c,d}); ref.~\cite{Dubouchet19}, (\textbf{e}). }
\label{Fig3} 
\end{figure*}

The presence of preformed electron pairs in a conventional low-$T_c$ superconductor with strong disorder was suggested by experiments of Sac\'{e}p\'{e} \textit{et al.}~\cite{Sacepe11} in scanning tunnelling measurements on a:InO films. A nearly full-width tunneling gap, $\Delta_{\text{PG}}$,  was visible at the temperature, where zero resistance develops, labeled $T=T_c$,  while a considerable density of states suppression was visible up to temperatures of $4T_c$.   Right below $T_c$ coherence peaks developed near the edges of the gap at a number of locations of the investigated area, while at other points no coherence peaks are found down to the lowest $T \ll T_c$. These observations, summarized in Fig.~\ref{Fig3}a and b, stimulated the  theory of the QBS, to be discussed below. Apart from the observations on a:InO similar pseudo-gap-type features were reported in subsequent experiments on a:InO, by Sherman et al\cite{Sherman12}, and on NbN, by Mondal et al\cite{Mondal11} , Chand et al\cite{Chand12}, and Carbillet et al. \cite{Carbillet16,Carbillet19}. In all cases, the observations are carried out on films with a $T_c$ strongly suppressed by disorder.  

In interpreting these experiments, it is reasonable to assume that a relatively large tunnelling gap is related to electron pairing, but it cannot serve as a direct measure of the superconducting coherence.  In addition,  the value is  too large in comparison to superconducting transition temperature $T_c$ to be understood by the BCS-type theory or its extension like the Eliashberg theory. Finally, it was found numerically~\cite{Ghosal01,Bouadim11} that the single-particle gap survives into the range of strong disorder, without the typical indications of superconducting coherence.  As an alternative interpretation of the data a suppression of the tunneling density-of-states near the Fermi-energy due to the dynamical Coulomb blockade~\cite{AAL80,LevitovShytov} has been suggested.  In low-$T_c$ disordered superconductors, it is experimentally not difficult to distinguish between dynamical Coulomb blockade and superconductivity-related pseudo-gap, due to the large difference in relevant energy scales: about 1 meV for pseudo-gap and tenths of meV for dynamical Coulomb blockade. An example of a joint analysis of dynamical Coulomb blockade together with the Finkel'stein mechanism of superconductivity-suppression has recently been discussed by Carbillet et al\cite{Carbillet19}. We believe that a theory based on preformed Cooper-pairs is the most plausible candidate. 

We proceed by addressing the formation of a collective gap for the condensation of the preformed Cooper pairs. Within the BCS theory, the collective gap  $\Delta_{\mathrm {col}} = 2\Delta_{\mathrm {BCS}}$, corresponds to the minimal excitation energy above the superconductive ground state, if single-particle excitations are forbidden.  Experimentally, as proposed by Deutscher\cite{Deutscher99,Deutscher05}, the collective gap of cuprate superconductors, $\Delta_{\mathrm {col}}$,  can be measured by point contact spectroscopy using Andreev reflection.  The voltage threshold for two-electron transfer, $2eV_{\mathrm {col}} = 2\Delta_{\mathrm {col}} $, which coincides with the threshold for single-electron tunnelling  $e V = \Delta_{\mathrm{BCS}}$, if the BCS relation is valid. 

A similar idea was implemented recently by Dubouchet \textit{et al.}~\cite{Dubouchet19} for amorphous InO close to the QBS. The point-contact differential  conductance $dI/dV$ was measured starting from the purely tunneling regime to a highly transmissive regime with a contact resistance of a few $k\Omega$, at which both single-electron and Andreev processes are relevant. In the latter regime additional peaks in $dI/dV$ emerged (see Fig.~\ref{Fig3}e) at  low voltages, $eV_{\mathrm {col}} \ll \Delta_{\mathrm{PG}}$, and \textit{in the superconducting state only}. Moreover, $V_{\mathrm{col}}$ was found to be weakly dependent on the tip location, but strongly temperature-dependent, similar to the $ \Delta_{\mathrm{BCS}}(T)$ dependence. Both features are in contrast with the behavior of the  single-particle threshold  $\Delta_{\mathrm{PG}}$. The magnitude  of $\Delta_{\mathrm{col}} \equiv eV_{\mathrm {col}} $ was interpreted as a genuine \textit{collective gap}, which develops together with the macroscopic superconducting coherence. This type of  behaviour was found only for strongly disordered a:InO films. In contrast, less disordered films (with $T_c \geq 3$K ) demonstrate single-gap BCS-type behaviour without any pseudo-gap feature.

These observations are analyzed by invoking the notion of the "parity gap", $\Delta_{\mathrm{par}}$, introduced in the seminal paper of Matveev and Larkin~\cite{Matveev97}, who developed the theory for an ultra-small superconducting grain. For sufficiently small grains the single-electron level spacing $\delta_1 = 1/\nu \mathcal{V} $ ($\nu $ the density-of-states per unit of volume and $\mathcal{V} $ the volume of the grain) exceeds the value of the superconducting gap $\Delta_{0}$  of the bulk of the material. The level spacing $\delta_1$ does not exceed the Debye energy. The condition $\delta_1 \gg \Delta_0 $ prevents the formation of many-body coherence, characteristic of the BCS ground-state, of the numerous single-electron states.  However two electrons, residing in the same localized orbital state with opposite spins, still attract each other and gain some energy $\Delta_{\mathrm{par}} \propto \delta_1$, with respect to the case when those electrons populate \textit{different} localized orbitals. 

We implement the idea of the parity gap to describe the pseudo-gap in  strongly disordered bulk (or 2D) materials following Ma and Lee.~\cite{Ma85}, Ghosal et al.~\cite{Ghosal01}, and Feigelman et al.~\cite{Feigelman07,Feigelman10a}. We assume, going beyond the theory described in Section IV, that the Fermi level for non-interacting electrons lies in the localized part of the band, but close to the mobility edge $E_c$. The  single-electron eigenfunctions are localized, with relatively long localization length $L_c \sim l \left[{(E_F - E_c)}/{E_c}\right]^{-\nu}$, where the exponent $\nu \approx 1.5$ in 3D. Then the matrix elements $P_i = \int d\mathbf{r} \psi_i^4(\mathbf{r})$ are nonzero in the thermodynamic limit and scale as $P_i \propto L_c^{-d_2} \sim \left[{(E_F - E_c)}/{E_c}\right]^{\nu d_2}$, due to the fractal nature of the electron eigenfunctions. The parity gap due to the local attraction between two electrons $\Delta_{\mathrm{par}} = {\lambda}/{\nu} \overline{ P_i}$ scales in the same way. This situation, similar to the one for ultra-small grains~\cite{Matveev97},  is realized when the single-electron spacing $\delta_1 = (\nu L_c^d)^{-1}$  far exceeds  the transition temperature $T_{c0}$ for the critical disorder ($L_c \to \infty$), see Eq.(\ref{Tc0}). Then it is possible to show~\cite{Feigelman10a} that the value of the parity gap $\Delta_{\mathrm{par}}$ lies between two other energy scales:
\begin{equation}
 T_{c0} \ll \Delta_{\mathrm{par}} \sim \delta_1 \left(\frac{T_{c0}}{\delta_1}\right)^\gamma \ll \delta_1
\label{DeltaP}
\end{equation}
When such a bulk disordered system with a tendency to electron pairing  does not develop superconducting coherence, the parity gap is seen both as a pseudo-gap in tunnelling experiments, and as an activation gap in transport. 

This physically rich theoretical model was developed to reach an understanding of the experimental data on a:InO\cite{Shahar92,Gantmakher96}. Large activation gaps, up to 10-15 K,  were reported in electronic transport measurements.  It was noted by Shahar and Ovadyahu\cite{Shahar92} that conventional scaling of the activation gap $\propto \delta_1 $ is excluded by the experimental data. As argued by Feigelman et al.~\cite{Feigelman10a}, the same data are consistent with a modified scaling of $\Delta_{\mathrm{par}}$ presented in Eq.(\ref{DeltaP}). Superconductivity will coexist with a parity gap when the ratio of $\delta_1/T_{c0}$ is not too large. The very presence of a solution with nonzero $T_c $ much smaller than both $\Delta_{\mathrm{par}}$ and $\delta_1$ is a new  feature of the theory of Ref.\cite{Feigelman10a} not anticipated in the original approach by Ma and Lee~\cite{Ma85}. It  is due to two effects that enhance the overlap matrix elements $M_{ij}$: the fractal nature of the eigenfunctions, and the Mott resonances~\cite{Mott71} between localized eigenstates with a small energy difference $ \omega \ll ~\delta_1$. 
The collective gap $\Delta_{\mathrm{col}}(T)$ at $T \ll T_c$ appears to be of the same order as $T_c$, and much less than $\Delta_{\mathrm{par}}$. An additional feature expected for the pseudo-gapped superconductors is the violation of the usual BCS rule that the full optical spectral weight is insensitive to superconducting transition~\cite{Feigelman10a}. Upon a further increase of disorder and of the level spacing $\delta_1$, the transition temperature $T_c$ and the collective gap $\Delta_{\mathrm col}$ gets smaller and eventually vanish, while $\Delta_{\mathrm{PG}}$ stays nonzero~\cite{Feigelman10a,Bouadim11}.  The resulting ground state is an insulator with preformed electron pairs.

For a system with preformed electron pairs, close to the QBS, the values for both $ T_c$ and $\Delta_{\mathrm{col}}$ are much smaller than the pseudogap $\Delta_{\mathrm{PG}}$. Therefore, this transition can be understood in terms of the Anderson pseudo-spin model~\cite{Anderson58}, which describes hopping of the preformed pairs between different localized orbitals $\psi_i(\mathbf{r})$. This specific case of the QBS can be called a "pseudo-spin" QBS, for short a pseudo-spin QBS. The hopping terms in the effective Hamiltonian are given by $H_{hop} = - \sum_{ij} J_{ij} (S_i^+ S_j^- + h.c.)$ where $S_i^{\pm}$ are the creation/annihilation operators of a preformed pair in the $i^{th}$ orbital. These hopping terms compete with the random on-site potential energy given by $H_{loc} = \sum_i 2\xi_i S^z_i$. The preformed pairs are just  hard-core bosons, and hence the operator set $S^z_i,S^+_i,S^-_i$ is formally equivalent to the spin-$\frac12$ operators $\mathbf{S}_i$. The superconducting state is then described by the non-vanishing quantum-statistical average values like $\Delta_i =  \langle S_i^- \rangle $, while in the insulating state all $\Delta_i \equiv 0$ and the preformed pairs are localized.

This pseudo-spin QBS appears to be in the same universality class as the order-disorder transition in the quantum XY spin-$\frac12$  model with random transverse fields. The transition is controlled by the value of the effective coupling strength $J \sim Z J_{ij}$, with $Z$ the typical number of connections per "spin". The theory worked out by Mézard, Ioffe and Feigelman\cite{Feigelman10b} leads to the following conclusions: 
\begin{itemize}
\item  a $T=0$ transition between superconducting and insulating states occurs at some critical value of the coupling strength $J_c$. 
\item at $J > J_c$, the superconducting state exists below a critical temperature $T_c(J)$, whose magnitude  drops very sharply to zero when $ J \to J_c$.
\item in a wide range of  $J > J_c$, the ordered state is extremely inhomogeneous, with a very broad probability distribution $\mathcal{P}(\Delta_i)$. Subsequent numerical studies for a 2D model by Lemarié et al\cite{Lemarie13} have confirmed this result.
\item a typical value of the order parameter $\Delta_{\mathrm typ} = \exp({\overline {\ln(\Delta_i)}})$  demonstrates an unusual exponential scaling near the $T=0$ critical point: $- \ln\Delta_{\mathrm typ} \propto (J - J_c)^{-1}$. This type of scaling was found earlier by Carpentier and Le Doussal\cite{LeDoussal00} for the disorder-driven Berezinskii-Kosterlitz-Thouless  transition.
\end{itemize}

We like to draw the attention to several unusual and important predictions of this pseudo-spin QBS. First, the inhomogeneity  of all superconducting properties close to the pseudo-spin QBS is much stronger than the emergent granularity in the bulk of the superconducting region. Moreover, the former develops at much larger spatial scale, which diverges close to the QBS. Eventually, the pseudo-spin QBS acquires superconducting features which are reminiscent of the classical percolation transition. This aspect might provide a framework to understand  the very unexpected size effects found near the QBS by Kowal et al\cite{Kowal08}, as well the unusual behavior of the Nernst coefficient at $T \gg T_c$ in near-critical a:InO films~\cite{Spathis08,Pourret09}.

The described mechanism for the pseudo-spin QBS  has one common feature with the "bosonic" scenario.  In both cases the relevant degrees of freedom are bosons (spin $1/2$ can be considered as bosons, but  with hard-core repulsion). The crucial physical difference is that  the pseudo-spin scenario does not assume  any local superconducting order with many-body correlations \textit{unless} global superconductivity has developed.  In particular, no trace of the coherence peaks is expected in the insulating state, in contrast to the standard bosonic scenario based on the Josephson-tunnel-junction model. The superconducting grain of a size large enough to have $\delta_1 \ll \Delta$ would show (smeared) coherence peaks in the tunnelling conductance on either side of the QBS. 

Another very important feature of the pseudo-spin scenario for the QBS  is that disorder plays a crucial role, unlike the commonly used Coulomb-blockade Josephson-tunnel-junction model. The pseudo-spin QBS model we summarize here is in that sense  similar to the  "infinite-disorder renormalization group" theory~\cite{Refael13,Igloi14}. The strong disorder effects and the need to account for the whole probability distribution of the relevant variable~\cite{Fisher95,LeDoussal00} make the use of an elementary scaling analysis problematic near such phase transitions. This might be the reason for the  broad range of  "critical exponents"  found in the literature.

%%%%%%%%%%%%%%%%%%%%%%%%%%%%%%%%%%%%%%
%%%%%%%%%%%%%%%%%%%%%%%%%%%%%%%%%%%%%%
%%%%%%%%%%%%%%%%%%%%%%%%%%%%%%%%%%%%%%
\vspace{0.5cm}
\section{Mesoscopic approach to inhomogeneous superconducting materials} 

In the past decades arrays made of Josephson-coupled networks based on superconducting islands coupled by tunnel barriers  have been studied~\cite{Fazio2001} extensively. The emphasis was on an analysis based on the Berezinskii-Kosterlitz-Thouless, the competition between Coulomb blockade and Josephson coupling, and the magnetic field dependence in view of the number of flux quanta per closed loop of the network. They have been also used as model-systems, to understand the QBS in real materials. From this perspective these tunnel-junction based systems ignore two important ingredients, relevant for real materials.  One ingredient is the use of large superconducting islands with a well-defined macroscopic quantum phase. In practice, much smaller pockets of superconductivity with a small number of electrons with a poorly defined phase may occur. Secondly, the coupling between the islands might in practice be much more transmissive than for a tunnel-barrier, either because of a few transmissive quantum channels or by a diffusive proximity-effect. Such type of arrays have been pioneered several decades ago \cite{Abraham1982}, and were recently addressed, with better lithography, in the work by Eley \textit{et al.}~\cite{Eley12}. In the latter work a  series of samples was studied with different coupling strengths between the islands, due to a variation in distance between the superconducting islands. The analysis was, as in the earlier work, based on the thermal Berezinskii-Kosterlitz-Thouless phase transition, assuming a Ginzburg-Landau proximity-effect description, with the coupling-strength varying exponentially with distance.

\begin{figure}[h]
   	\includegraphics[width=1\linewidth]{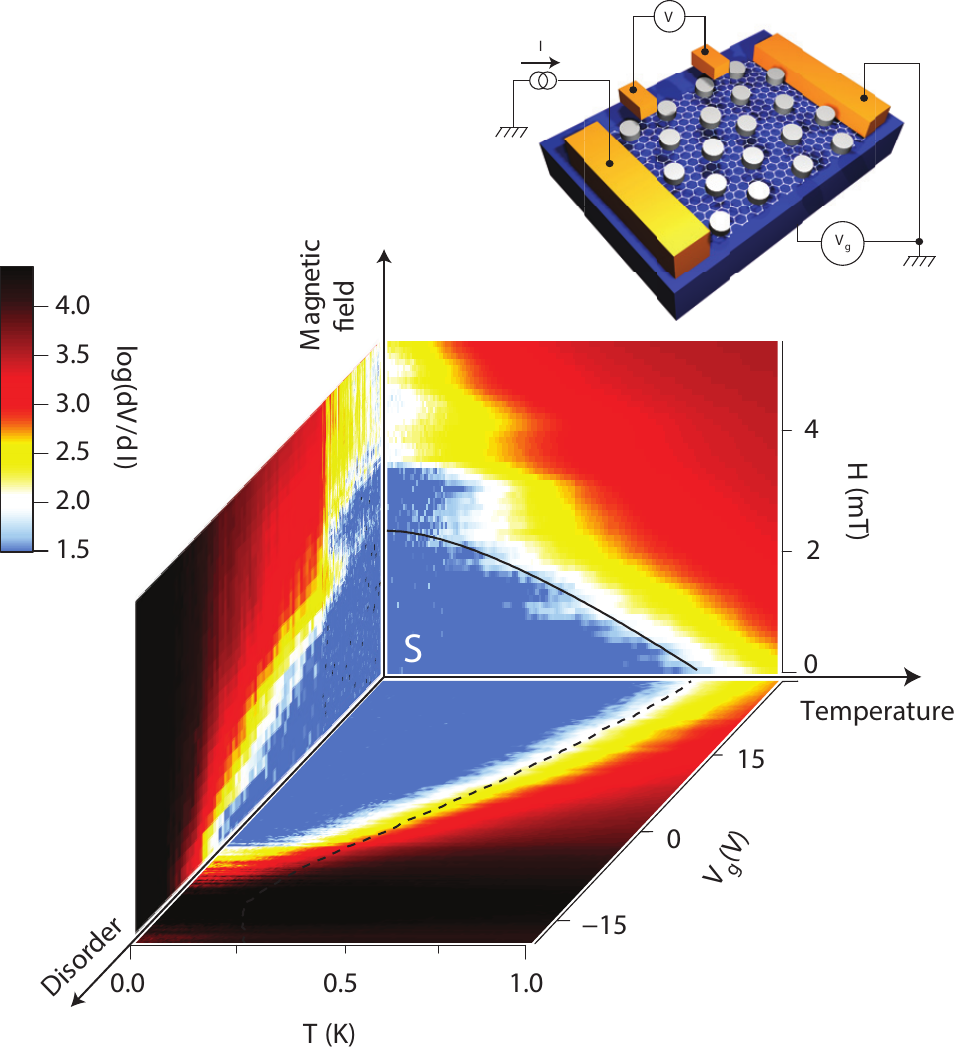} 		
\caption{\textbf{Quantum breakdown of superconductivity in a mesoscopic device.} A 3D-phase diagram showing the superconducting state reconstructed from measurements of the array resistance in back-gate voltage, $V_{\text{g}}$, magnetic field, $H$, and temperature, $T$,  space~\cite{Han14}. The resistance in ($V_{\text{g}}$ , $H$) space is measured at 0.06~K. Notice that traces of a re-entrance of superconductivity above the first critical field are visible in both the ($V_{\text{g}}$ , $H$) and ($V_{\text{g}}$ , $T$) planes,  a clear signature of the mesoscopic superconductivity~\cite{SpivakZhou95}. The inset shows a schematics of the mesoscopic sample: a graphene Hall bar decorated with an array of superconducting discs. Reproduced from ref.~\cite{Han14}.}
\label{Fig.5} 
\end{figure}

The discovery of tunable 2-dimensional materials such as graphene in combination with superconductors makes it possible to construct a Josephson-arrays with an \textit{in situ} change of coupling strength. This has been realized in a recent experiment by Han \textit{et al.}~\cite{Han14}. A layer of graphene was decorated with a triangular array of circular tin disks, each with a diameter of 400 nm and a mutual distance between the centers of the disks of 1 micrometer. The total number of disks covered an area of 5 by 10 micrometer. The sample was equipped with a gate which allowed the tuning of the carrier-density in the graphene, which meant that the Josephson-coupling could be tuned. The conductivity of the array could be tuned from a full superconducting state through an 'anomalous metal' phase, going over to an approximate insulating phase, with diffusive transport in the graphene. The data as a function of gate voltage for different magnetic fields and temperatures clearly resemble (Fig.\ref{Fig.5}) the phase diagram laid out by M. Fisher~\cite{Fisher90}. These results are already interesting and important as such, showing the importance of quantum phase fluctuations in Berezinskii-Kosterlitz-Thouless-physics,  but the experimental method allows for several possible further routes to be explored. In principle, once could consider smaller islands in order to get the islands with an energy-spacing for the electrons larger than the superconducting energy gap, although this might require the use of superconductors with a low carrier density. A second route could be the use of ballistic graphene, which is in principle possible by using BN-encapsulated graphene. Thirdly, the presently small scale of the array, about 20 sites, may lead to a contribution from finite-size effects on the observed quantum Berezinskii-Kosterlitz-Thouless-like transition, which need to be evaluated by making arrays with more sites.   

B\o{}ttcher et al\cite{Bottcher18} carried out a similar experiment using a rectangular array of square-shaped superconducting aluminium islands deposited on a two-dimensional tunable 2-dimensional electron gas in the semiconductor InGaAs-system, using the materials systems developed for Majorana-physics. The published results are analogous to data presented in earlier work on other 2 D arrays and analyzed by the conventional scaling analysis~~\cite{Fisher90}. In principle this experimental system has the potential to benefit from insights in mesoscopic superconductivity. The Andreev-reflection process which mediates the phase-coupling between the superconducting islands has been thoroughly analyzed by Kjaergaard \textit{et al.}\cite{Kjaergaard2017}. The critical new step is to integrate the understanding of the proximity-effect from the perspective of Andreev-processes at the interface between the normal metal and the superconductor. This knowledge is well-developed and continues to be tested in various hybrid systems. The important ingredient of the study of Josephson-coupled arrays is the fate of the quantum-phase of each superconducting island. Further research, with considerably smaller superconducting islands appears to be within reach. 
     
The emphasis on the tunability of the Josephson-coupling energy in controlling the macroscopic transport properties of the arrays calls into question the dependence on environmental noise. Martinis and co-workers~\cite{Martinis87,Martinis93} have addressed this subject in the context of macroscopic quantum tunneling. Additionally, the thermal blackbody-radiation has been found to contribute significantly  to the performance of superconducting quantum circuits~\cite{Barends2011} A reminder of the importance of the sensitivity to environmental radiation for materials-research was shown in recent experiments by Tamir \textit{et al.}~\cite{Tamir19}, which showed it in the resistive transition of a:InO, but also of a crystalline 2D superconductor, H2-NbSe$_2$. A similar dependence was found recently by Dutta et al.~\cite{Dutta19} in the  onset of resistance in the vortex state of a-MoGe-films. It is a reminder that experiments on 'weak' superconductors, on the verge of quantum breakdown, need to be carried out in an electromagnetically well-shielded cryogenic environment, common for mesoscopic research but not typical for materials research.

%%%%%%%%%%%%%%%%%%%%%%%%%%%%%%%%%%%%%%
%%%%%%%%%%%%%%%%%%%%%%%%%%%%%%%%%%%%%%
%%%%%%%%%%%%%%%%%%%%%%%%%%%%%%%%%%%%%%
\vspace{0.5cm}
\section{Applications to quantum circuits and (protected) qubits} 
   
 One of the most interesting and fascinating potential applications of strongly disordered superconductors is the superinductor. A superinductor is a  non-dissipative element of an electrical circuit with an impedance $Z=\sqrt{(L/C)}$ much larger than the resistance quantum $R_Q = h/{(2e)^2}$. The inductance of a superconducting wire consists of the geometric inductance, which stores the energy of the electromagnetic field, and the kinetic inductance of the superconducting condensate, which stores the kinetic energy of the moving superfluid. The geometric inductance scales with the length of the wire and depends only logarithmically on the diameter of the wire. Its value is typically $1~pH/{\mu m}$, which makes the geometric inductance unusable, a problem well known in for example RF electronics\cite{Kang2018}. The alternative is to exploit the kinetic inductance of a low-dimensional material such as graphene\cite{Kang2018}, which for normal conduction leads to high losses and hence low quality factor if used in a resonator. The low loss can be provided by a  superconductor of which the kinetic inductance is in general given by: $L_k= \hbar R_n/{\pi \Delta}$, which means that a high normal state resistivity and a low superfluid-density can be used to maximize the kinetic inductance. Ideally, one would like to achieve at least an inductance per unit length of $0.1~nH/{\mu m}$, which means films with $0.5-2.5~nH/\square$. With such a superinductor one can construct protected qubits reviewed in Doucot and Ioffe~\cite{Doucot2012}, with new ideas put forward by Brooks \textit{et al.}~\cite{Kitaev2013}, Groszkowski \textit{et al.}~\cite{Groszkowski2018} and Smith \textit{et al.}~\cite{Smith2019}.      

\begin{figure}[t!]
   	\includegraphics[width=1\linewidth]{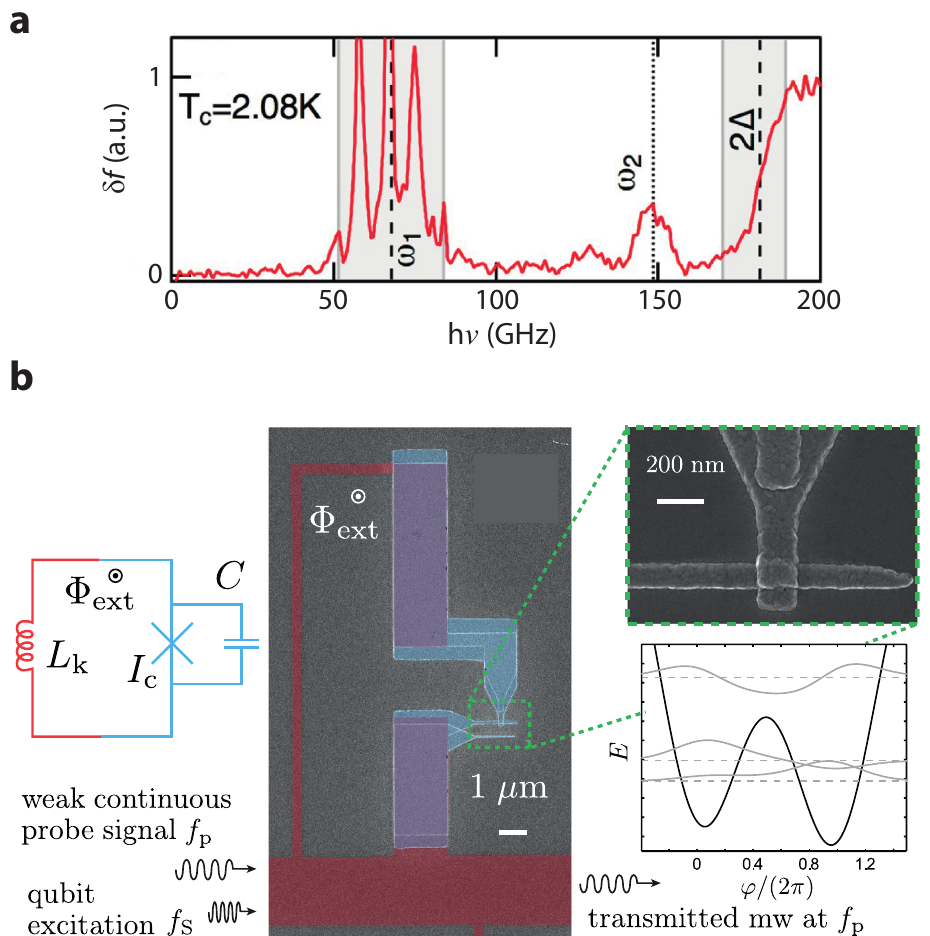} 		
		    	\caption{\textbf{Two examples of use of high microwave kinetic inductance materials: spectroscopy and quantum circuits.} \textbf{a,} Typical absorption spectrum of a superconducting resonator made from granular aluminum in use for astronomical detectors and as superinductor, showing the conventional onset of absorption at $2\Delta$ and subgap resonances~\cite{Grenet19}, identified as two distinct groups of absorption lines labeled as $\omega_1$ and $\omega_2$. These modes are most likely collective modes coupled to the phase-differences. The signal is the shift of the resonant frequency in the 2 to 6 GHz range due to the absorption of the radiation in the frequency range from 10 to 200 GHz. \textbf{b,} Hybrid RF SQUID used as a qubit and consisting of an Al SIS superconducting tunnel-junction, shunted by a high kinetic inductance loop made of TiN. Reproduced from ref.~\cite{Grenet19}, APS (\textbf{a}); ref.~\cite{Peltonen2018}, (\textbf{b}).}
   	\label{Fig4}
   \end{figure}

Motivated by these ideas about protected qubits and another proposal put forth by Mooij, Nazarov and Harmans\cite{MooijHarmans2005,MooijNazarov2006} on quantum phase slip junctions Astafiev and co-workers have studied the quantum coherence of superconducting nanowires for a range of materials like a:InO\cite{Astafiev12}, NbN\cite{Peltonen2016}, and TiN\cite{Peltonen2018,DeGraafAstafiev2019}. Rather than a study of thermally activated phase slips processes, coherent quantum phase slip processes are studied. The critical quantity is the transition amplitude for quantum phase slips, expressed in energy as $E_s$, reaching values in the 100 GHz range. In addition, one needs an impedance of the environment larger than the quantum unit of resistance of 6.45 $k\Omega$. The scale of the inductive energy $E_L$ is given by ${\Phi_0}^2/{2L}$ with $L$ the kinetic inductance of the wire.  In order to reach $E_s>>E_L$ one needs superconducting materials with a high normal state resistivity, \textit{i.e.} the strongly disordered superconductors, which are central in this review (Fig.~\ref{Fig4}b). 

An alternative experimental strategy is the work of Kuzmin \textit{et al.}~\cite{Kuzmin18}. Instead of a model-system to study the breakdown of superconductivity they constructed a model-system in which they could experimentally focus on the collective electromagnetic phase-mode. They created a linear chain of 40.000 Josephson-tunnel-junctions and used the device to study the collective modes of microwave photons. These modes have a velocity as low as $v \approx 10^6$ m/s and a wave impedance as high as $Z\approx R_Q$, with $R_Q=h/{(2e)}^2\approx 6.5~k\Omega$, the superconducting resistance quantum. A superconductor-insulator transition is obtained by changing the junction-area~\cite{Kuzmin18}, which leads with a smaller area to the quantisation of the charge on the superconducting island, and hence to strong fluctuations of the quantum-phase on the islands, visible, for example, as a breakdown of the collective mode.   

In the search for suitable materials, thin films of granular aluminium (GrAl) are being explored as a possible candidate~\cite{Grunhaupt2018,Grunhaupt2018PRL,Grunhaupt2019}. This subclass of materials has been known for a long time to be easily accessible, but at the same time difficult to understand. The material is made by deposition of aluminium in a partial pressure of oxygen leading to an assembly of small aluminium grains, surrounded by an AlO$_x$ layer, hence the name 'granular'.  For increasing resistivity the critical temperature increases up to a large factor of 3 to 4, an observation, which has so far defied a coherent explanation. A number of promising properties have been measured, at the same time some detrimental properties have also been identified, including the microwave properties~\cite{Grenet19}, Fig.~\ref{Fig4}a. The same material is also studied actively by Gershenson and co-workers~\cite{Gershenson2019a,Gershenson2019b}.  The parameter-details for granular aluminium are not very well known. It is possible that they can be considered as random arrays of tunnel-junctions, small aluminium particles surrounded by an oxide tunnel-barrier, which connects the aluminium grains. The difference is however, that compared to the experiment by Kuzmin \textit{et al.}~\cite{Kuzmin18}, as well as in the earlier work of Fazio\textit{ et al.}~\cite{Fazio2001}, the aluminium islands in granular aluminum are very small.  The size of the grains may lead to a situation in which energy level spacing exceeds the pairing gap, which may be important for the understanding of the results reported by Lévy-Bertrand \textbf{et al.}~\cite{Grenet19}.   

The experimental work described in the previous paragraphs makes clear that an ideal superinductor does require thoughtful experimental work. At the same time the research on the QBS has uncovered the various ways in which superconductors with high normal state resistivity loose their superconducting properties. These pathways are taking into account the nature of the coupling of the superconducting phases, as well as the possibility of localized pairs  and the pairing strength.  With the ultimate goal of lossless inductors the materials-requirements have been discussed in recent publications by Feigel'man and Ioffe\cite{Feigelman15,Feigelman2018}. Many challenges towards a usable superinductor are ahead of us and will need to be compared with the use of arrays of tunnel-junctions aiming for the same functionality\cite{Kuzmin18}.

%%%%%%%%%%%%%%%%%%%%%%%%%%%%%%%%%%%%%%%
%%%%%%%%%%%%%%%%%%%%%%%%%%%%%%%%%%%%%%%
%%%%%%%%%%%%%%%%%%%%%%%%%%%%%%%%%%%%%%%
\vspace{0.5cm}
\section{Open problems and conclusions}

 For superconductors the challenge is that a \textbf{macroscopic} \textit{quantum} state emerges from or gets destroyed by \textbf{microscopic} \textit{quantum} properties. In this review we have identified several pathways, which may occur in real materials. Since our interest is not so much in one specific model system, nor in just plausible interpretations of experiments, but in a description of real world materials, the additional challenge is to identify,  which pathway is most likely applicable to a specific material-system, while at the same time not ruling out that not all possible pathways have already been identified or properly solved. 

We point out that several experimental observations have not yet been understood:  
\begin{itemize}
\item For homogeneously disordered superconductors, assessing an effective Coulomb interaction is conceptually difficult: The absence of grain boundaries makes undefined an equivalent capacitance that determines charging effects in tunnel junctions and the ensuing phase fluctuations. Addressing quantum phase fluctuations in superconductors on the basis of Josephson-junction models remains therefore purely qualitative and should not disguise the fact that exact role of Coulomb interaction on phase fluctuations in disordered superconductors is an open question.
\item   
In the vicinity of a superconductor-to-metal transition, a strange metal state" is frequently observed~\cite{KKS19}.  Do these observations indicate the existence of an intrinsic equilibrium  quantum state, which may result from the intricate interplay between Cooper pairing, Coulomb repulsion and Anderson localization?  Or are these observations in need of a careful reexamination of the experimental conditions under which the data have been taken, because  of a possible coupling to the non-equilibrium environment? 
\item The experimentally observed high sensitivity~\cite{Tamir19,Dutta19} to radio-wave interference, as known for Josephson-junctions of SIS and SNS-type~\cite{Martinis93}, calls for an analysis of the effect of microwaves on the study of superconducting materials, which are expected to be spatially inhomogeneous. Both the common Josephson-response due to a local voltage difference, affecting the local phase-differences,  should be considered as well as the non-equilibrium effects, which change the Josephson-coupling.  
\item For the magnetic-field-driven QBS in moderately disordered materials the nature of the strongly pinned vortex glass in the $T=0$ limit is still unclear~\cite{Sacepe19}. The theoretical problem is to describe a vortex glass with a macroscopic superfluid density $\rho_s$. In the strong disorder limit, a combined account of the localization of preformed pairs and Coulomb interaction is still to be developed.
\item Low energy collective modes in pseudogap superconductors: can one avoid them? This is an issue of practical importance for the development of various quantum circuits where a high kinetic inductance $\propto 1/\rho_s$ should be combined with an  absence of dissipation at frequencies in the sub-GHz range.
\item The giant magnetoresistance peak terminating superconductivity in some materials (a:InO~\cite{gantmakher2000,Sambandamurthy04,Steiner05}, TiN~\cite{Baturina07a}, nano-patterned a-Bi~\cite{Stewart07,Nguyen09}, proximitized graphene~\cite{Allain12}) continues to defy understanding. New experimental approaches must be employed to probe this insulating state beyond transport measurements that have proven to be limited by non-equilibrium effects~\cite{Doron17}.
\item On the insulating side of a superconductor-to-insulator transition we may encounter a many-body localized state, with extremely weak electron-phonon coupling~\cite{Ovadia09} and conductivity vanishing at some nonzero temperature~\cite{Ovadia15}. 
\item Lastly, understanding which microscopic parameters and/or phenomena define the metallic-like or insulating nature of the ground state terminating superconductivity for given material, that is, SMT versus SIT, remains a major question.
\end{itemize}

Future research is expected to focus on the interaction between the three themes addressed in this review: theory, experiments and applications. Experiments unavoidably will have to focus on model-systems, which allow them to evaluate identified theoretical pathways. Applications have the virtue to provide large amounts of experimental data, assuming they have been taken under relevant experimental circumstances, knowing the vulnerability of the system to the environment. From the latter point of view the increased use of strongly disordered superconducting materials for astronomical instrumentation and quantum computation holds the potential of providing data, which are significant for the evaluation based on the available theoretical models. 

Finally, it is to be expected that the subject of quantum breakdown of superconductivity will benefit from the increasing availability of gatable 2-dimensional superconducting systems~\cite{Caviglia08,Tsen16,Cao18}, which constitute well-defined atomically uniform model systems in which specific pathways could be tested in much detail.

\section*{Acknowledgments}
The authors like to thank the participants of the workshop on \textit{The challenge of 2-dimensional superconductivity} (July 8- 12, 2019, Lorentz Center, University of Leiden) for providing us with up to date insight into the various viewpoints on the subject. B.S. has received funding from the European Research Council (ERC) under the H2020 programme (grant No. 637815) and from the French National Research Agency (ANR grant \textit{CP-Insulator}). M.F. is supported by a Skoltech NGP grant and by the RAS program "Advanced problems in low temperature physics". T.M.K. is supported by a grant from the Russian Science Foundation (No. 17-72-30036) and by the Würzburg-Dresden Center of Excellence on \textit{Complexity and topology in quantum matter} (CT.QMAT).
\vspace{1cm}

%\textbf{Competing Interests} The authors declare that they have no competing financial interests.

%\bibliography{SIT_review_bib1}

\end{document}